# Effect of the benzyl groups on the binding of $H_2$ by three-coordinated Ti complexes

**Hoonkyung Lee**

*Department of Physics, University of California, Berkeley, California 94720, USA*

*Materials Science Division, Lawrence Berkeley National Laboratory, Berkeley, California 94720, USA*

Email: hkiee3@civet.berkeley.edu

ABSTRACT

Using first-principles calculations, we investigate the adsorption of $H_2$ molecules on a three-coordinated benzyl-decorated titanium complex suggested in a recent experiment [Hamaed *et al*., J. Am. Chem. Soc. 130, 6992 (2008)]. Unlike the interpretation of the experimental results that the Ti(III) complex can bind five $H_2$ molecules via the Kubas interaction, the Ti(III) complex cannot adsorb $H_2$ molecules via the Kubas interaction. In contrast, a benzyl-released Ti(III) complex can adsorb up to two $H_2$ molecules with a binding energy of ~0.25 eV/$H_2$ via the Kubas interaction, in good agreement with the measurement of ~0.2 eV. The calculated occupation number of $H_2$ molecules at 25 $^o$C and −78 $^o$C under 60 atm is 0.9 and 1.9, respectively, in good agreement with the measurement of 1.1 and 2.4 near the conditions, respectively. Our results suggest that the Ti complex in experiment might be a benzyl-released form.

In the last decade, hydrogen storage in various metals or chemical hydride materials has been intensively studied, and for some materials the hydrogen storage capacity has been substantial.[1,2] However, there are problems such as high dehydrogenation temperatures, slow kinetics, and poor reversibility.[1,2] For these reasons, nanostructured materials adsorbing hydrogen on their surfaces have been considered as potential hydrogen storage systems because of the possibility of achieving low desorption temperatures, fast kinetics, and good reversibility.[3,4] However, it has been found that the hydrogen storage capacity of nanomaterials sharply decreases at room temperature and ambient pressure[5,6] because of the small coupling between hydrogen and the storage materials mediated by the van der Waals interaction (~0.04 eV).

In recent years, in order to enhance the interaction of $H_2$ and the storage system, theoretical studies have proposed that the Kubas interaction[7] will enable transition metal atoms to attract $H_2$ molecules through the hybridization of the $d$ states with the $H_2$ states.[8,9] Some theoretical studies have shown that transition metal (i.e., Sc, Ti, and V)-decorated nanostructured materials[8-14] can bind several $H_2$ molecules per metal atom with a desirable binding energy of ~0.2 – 0.6 eV for reversible hydrogen storage at ambient conditions[15], and also satisfy the goal of 9 wt % set by the Department of Energy (DOE) by the year 2015[16].

On the experimental side, there have been efforts[17,18] devoted to synthesizing hydrogen storage materials employing the Kubas interaction, namely, chemically reducible Ti oxides and Ti-ethylene complexes. However, the interaction of $H_2$ molecules in these systems is still short of the expected strength. More recently, an enhanced interaction with the binding energy of $H_2$ molecules ~0.2 eV on three-coordinated benzyl-decorated Ti(III) complexes attached to a silica surface has been observed[19], which corresponds to the binding energy range by the Kubas interaction estimated by previous theoretical results[8-10]. Hamaed et al. have interpreted that the benzyl-decorated Ti(III) complex can adsorb up to five $H_2$ molecules via the Kubas interaction. This result appears to be a promising step toward the possibility of hydrogen storage on metal-decorated nanostructured materials.

In this paper, we investigate the question of whether $H_2$ molecules bind to the benzyl-decorated

Ti(III) complex via the Kubas interaction, and compare the calculated occupation number of $H_2$ molecules with the experimental measurement. We find that contrary to the proposed interpretation of recent experimental results, the benzyl-decorated Ti(III) complex cannot adsorb $H_2$ molecules via the Kubas interaction. In contrast, a benzyl-released Ti(III) complex can bind up to two $H_2$ molecules with a binding energy of ~0.25 eV/$H_2$ via the Kubas interaction, which is good agreement with the measured value of ~0.2 eV[19]. Another attractive feature is that this structure is more stable than the benzyl-decorated Ti(III) complex proposed in the experiment[19]. In addition, the calculated occupation number of $H_2$ molecules on the benzyl-released Ti(III) complex is in good agreement with the measurements[19]. Our results show that the three-coordinated Ti complex in experiment might be a benzyl-released form.

Our calculations were carried out using first-principles density functional calculations with a plane-wave-based total energy minimization[20]. The exchange-correlation energy functional of the generalized gradient approximation (GGA) of Perdew, Burke, and Ernzerhof was used,[21] and the kinetic energy cutoff was taken to be 35 Ryd. The optimized atomic positions were relaxed until the Hellmann-Feynman force on each atom was less than 0.01 eV/Å. Supercell[22] calculations throughout were employed where the adjacent nanostrucutres were separated by over 10 Å.

To order to investigate the binding mechanism of $H_2$ molecules on the Ti(III) complex based on the suggested structure on a silica surface in the experiment[19], we construct a model for the benzyl-decorated Ti(III) complex as shown in Fig. 1(a) where the Ti atom is bonded with a benzyl group ($-CH_2Ph$; Ph indicates a phenyl group, $-C_6H_5$) and with each of two oxygen atoms bonded with $SiH_3$ and the Si-Si distance in the optimized geometry is calculated to be 5.3 Å. We find that the phenyl group of the benzyl group is strongly bonded with the Ti atom through the hybridization of the Ti $d$ states with the Ph $\pi$ or $\pi^*$ states (the so-called Dewar interaction[7]) as shown in Figs. 1(a) and 1(b). The distance between the Ti atom and the nearest carbon atom is 2.4 Å. Only the difference between our model and the Ti(III) complex on a silica surface may be the Si-Si distance, which might affect the binding of $H_2$ molecules by the interaction change between the Ti atom and the phenyl group. We have confirmed that the adsorption of H$_2$ molecules on the Ti(III) atom is not affected regardless of the Si-Si distance. Therefore, our model gives a good description of the adsorption of H$_2$ molecules on Ti(III) complex attached to a silica surface as explored in Ref. 19 as long as the three coordination of the Ti atom on the silica surface is maintained. The molecular formula of our system is expressed as 2SiH$_3$·2O·Ti·CH$_2$Ph. Figure 1(b) shows that a H$_2$ molecule adsorbs on the Ti atom with a binding energy of ~0.04 eV, and the distance between the Ti atom and the H$_2$ molecule is 3.3 Å which corresponds to the equilibrium distance for the van der Waals interaction. These results show that H$_2$ molecules do not bind to the benzyl-decorated Ti(III) complex via the Kubas interaction. This is in contrast to the previous interpretation of the experimental results[19].

Next, we replace the benzyl group with a hydrogen atom as shown in Fig. 1(c) to examine how the interaction between the Ti atom and the benzyl group influences the adsorption of H$_2$ molecules. We find that the benzyl-released Ti(III) complex can bind up to two H$_2$ molecules with binding energies of 0.33 and 0.26 eV/H$_2$ for the first and second H$_2$ molecules, respectively, as shown in Fig. 1(d). The distance between the Ti atom and the H$_2$ molecules is ~2.0 Å, and the bond length of H$_2$ is slightly elongated to 0.79 Å from 0.75 Å in vacuum. This corresponds to the Kubas interaction described in the literature[8-11]. Therefore, the benzyl-decorated Ti(III) complex does not adsorb H$_2$ molecules via the Kubas interaction because of the interaction between the Ti atom and the benzyl group. Furthermore, the maximum number of attachable H$_2$ molecules on the Ti(III) atom does not agree with the estimated of five using the 18-electron rule in Ref. 19. In the case of the benzyl-decorated Ti(III) complex, the total number of electrons contributing to the Ti atom is 13 from 3 (the number of the Ti bondings with two oxygen atoms and one benzyl group's carbon atom) plus 4 (the number of the Ti valence electrons) plus 6 (the number of the pi electrons of the phenyl group) if all the pi electrons of the phenyl group contribute to the Ti atm. According to the 18-electron rule, up to two H$_2$ molecules should be adsorbed. However, the benzyl-decorated Ti (III) complex cannot adsorb H$_2$ molecules. Therefore, we conclude that this rule is not applicable to these systems made up of Ti(III) complexes although the rule has been successful in transition metal-olefin complexes[8,23].

To investigate the origin that the benzyl-decorated Ti(III) complex cannot adsorb $H_2$ molecules, we calculate the projected density of states (PDOS) for the benzyl-decorated or benzyl-released Ti(III) complexes for comparison as shown in Figs. 2(a) and 2(b), respectively. For the benzyl-decorated Ti(III) complex, the unoccupied Ti $3d$ states which are responsible for binding $H_2$ molecules remain unchanged compared to those in the benzyl-released Ti (III) complex even though the unoccupied Ti $3d$ states are hybridized with the phenyl group. This implies that the benzyl group may not affect significantly the Ti $3d$ states binding $H_2$ molecules. Therefore, the reason why the benzyl-decorated Ti(III) complex cannot bind $H_2$ molecules through the Kubas interaction might be attributed to steric hindrance for adsorption of the $H_2$ molecules due to the repulsive interaction between the $H_2$ molecules and the benzyl group.

To compare our results with the experimental measurements, we obtain the occupation number for $H_2$ molecules on a site in equilibrium between adsorbed $H_2$ molecules and $H_2$ gas (a reservoir). The grand partition function is given by $Z = \sum_{n=0}^{N_{max}} g_n e^{n(\mu-\varepsilon_n)/kT}$ for a multiple $H_2$ binding per a site (Ti) where $\mu$ is the chemical potential of the $H_2$ gas, $-\varepsilon_n$ (>0) and $g_n$ are the binding energy of the adsorbed $H_2$ molecules per $H_2$ and the degeneracy of the configuration for a given adsorption number of $H_2$ molecules $n$, respectively, $N_{max}$ is the maximum number of adsorbed $H_2$ molecules per site, $k$ and $T$ are Boltzmann constant and the temperature, respectively. From the relation $f = kT \partial \log Z / \partial \mu$, the fractional occupation number per site is:

$$f = \frac{\sum_{n=0} g_n n e^{n(\mu-\varepsilon_n)/kT}}{\sum_{n=0} g_n e^{n(\mu-\varepsilon_n)/kT}}. \qquad (1)$$

Figure 3 shows the occupation number of $H_2$ molecules on the benzyl-released Ti(III) complex as a function of the pressure and temperature. We used the experimental chemical potential of $H_2$ gas[24], and the energy ($-\varepsilon_n$) for the value reduced by 25% from the calculated binding energy of $H_2$ molecules because of the zero-point vibration energy[10]. The occupation number of $H_2$ molecules $f$ at 25 °C and 60 atm is 0.9, and that at −78 °C and 60 atm is 1.9 as shown in Fig. 3. This is attributed to the Gibbs factor ($e^{(\mu-\varepsilon_1)/kT}$) for the binding of one $H_2$ molecule which dominates at 25 °C and 60 atm ($\mu = -0.21$ eV, $\varepsilon_1 = -0.25$ eV, and $\varepsilon_2 = -0.20$ eV), and $e^{2(\mu-\varepsilon_2)/kT}$ for the binding of 2 $H_2$ molecules which dominates at −78 °C and 60 atm ($\mu = -0.10$ eV). These numbers are in good agreement with the measurement[19] of 1.1 $H_2$ and 2.4 $H_2$ molecules at the above conditions, respectively.

In order to investigate the stability of the benzyl-decorated Ti(III) complex and the benzyl-released Ti(III) complex, the formation energy for the complexes is calculated by:

$$F = E[\text{Ti-H}] + E[\text{Benzyl-H}] + \mu_{\text{Benzyl-H}} - E[\text{Ti-Benzyl}] - E[H_2] - \mu_{H_2} \quad (2)$$

where E[X] indicates the total energy of the systems for X, and Ti-H, Benzyl-H, and Ti-Benzyl stand for the benzyl-released Ti(III) complex, a hydrogen passivated benzyl group, and the benzyl-decorated Ti(III) complex, respectively, and $\mu_{\text{Benzyl-H}}$ and $\mu_{H_2}$ are the chemical potential of $H_2$ gas and a Benzyl-H phase, respectively. At the temperature of 180 °C when synthesizing the Ti(III) complex in the experiment[19] and the $H_2$ or Benzyl-H pressure of 1 atm, the formation energy is calculated to be −0.25 eV where $\mu_{H_2}$ (=−0.53 eV) and $\mu_{\text{Benzyl-H}}$ (=−1.18 eV) were used with the experimental values for $H_2$ gas and propane ($C_3H_8$) gas, respectively[24]. This result shows that the benzyl-released Ti(III) complex may be more stable than the benzyl-decorated Ti(III) complex suggested in the experiment[19]. The actual formation energy may be lower than −0.25 eV because the atomic mass of the $C_7H_8$ is greater than that of the $C_3H_8$ (i.e., $\mu_{\text{Benzyl-H}} < \mu_{\text{Propane}}$ from the chemical potential for ideal gas, $\mu_{\text{ideal}} \propto -\log^m$ where $m$ is the mass of a particle), so that the approximation for the chemical potential of the Benzyl-H does not affect the conclusion. We think that the three-coordinate Ti complex in experiment might be a benzyl-released form.

In summary, we have investigated the enhanced interaction of $H_2$ molecules on a three-coordinated Ti(III) complex via the Kubas interaction measured in recent experiment using first-principles calculations. The benzyl-decorated Ti(III) complex suggested in the experiment is not able to bind $H_2$ molecules through the Kubas interaction, which is contradictory to the experimental interpretation. In contrast, a benzyl-released Ti(III) complex can bind up to two $H_2$ molecules with an enhanced binding energy of ~0.25 eV through the Kubas interaction, which well agrees with the measured binding energy of ~0.2 eV. Furthermore, the calculated occupation number of $H_2$ molecules at a given temperature and pressure is in good agreement with the measurements. Therefore, we suggest that the Ti(III) complex on a silica surface in experiment might be the benzyl-released form.

This research was supported by the National Science Foundation Grant No. DMR07-05941 and by the Director, Office of Science, Office of Basic Energy Sciences, Materials Sciences and Engineering Division, U. S. Department of Energy under Contract No. DE-AC02-05CH11231. Computational resources were provided by NPACI and NERSC.

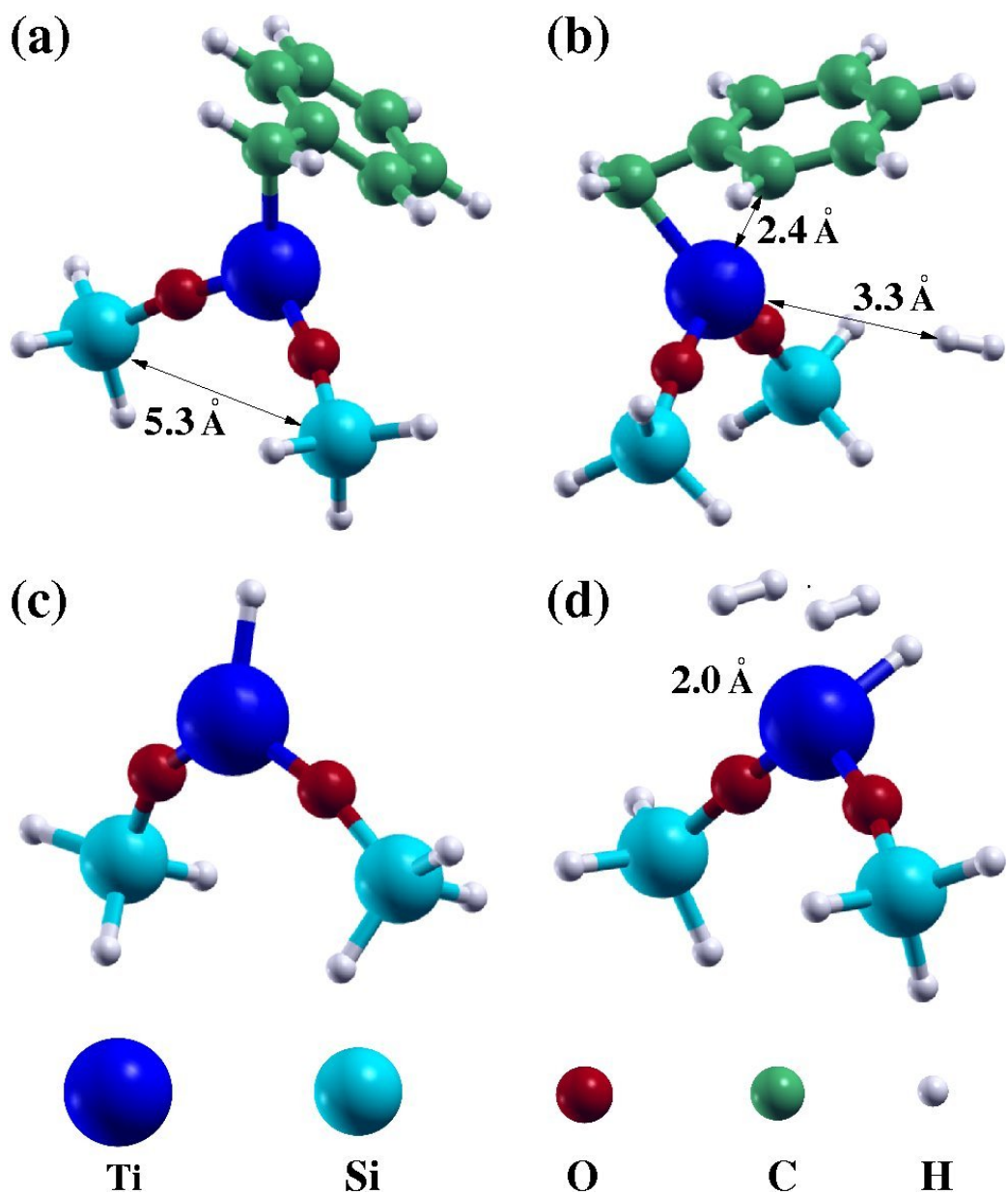

**Figure 1 (Color online):** Optimized atomic geometries of the three-coordinated Ti(III) atom and the attachment of H$_2$ molecules to the Ti atom. (a) and (b) show the optimized atomic geometries for a benzyl-decorated Ti(III) complex and the adsorption of H$_2$ molecule on the Ti(III) complex, respectively. (c) and (d) show the atomic geometries for a benzyl-released Ti(III) complex where the benzyl group in Fig. 1(a) is replaced by a hydrogen atom and the adsorption of H$_2$ molecules on the Ti(III) complex with the benzyl group replaced by a hydrogen atom, respectively.

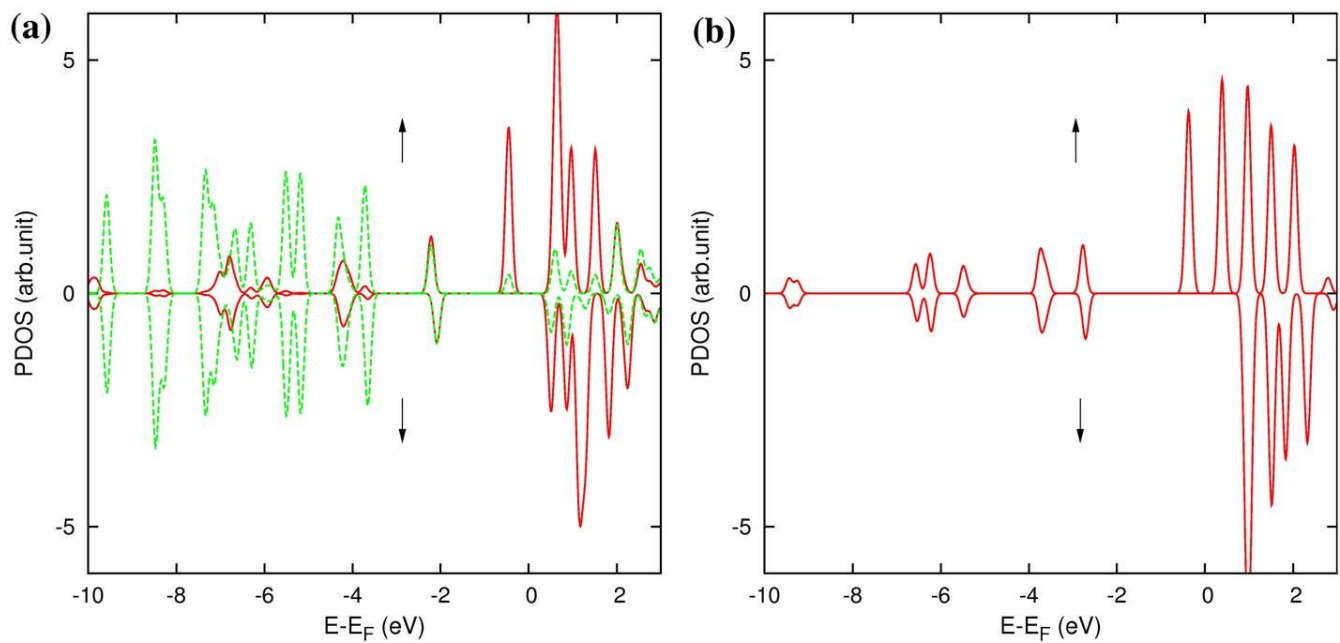

**Figure 2 (Color online):** (a) PDOS of the Ti 3$d$ orbitals and phenyl group of the benzyl-decorated Ti(III) presented in Fig. 1(a). (b) PDOS of the Ti 3$d$ orbitals of the benzyl-released Ti(III) presented in Fig. 1(c). The Fermi level is set to zero. The solid and dotted lines indicate the Ti 3$d$ states and the phenyl states, respectively. The arrows indicate majority and minority spins.

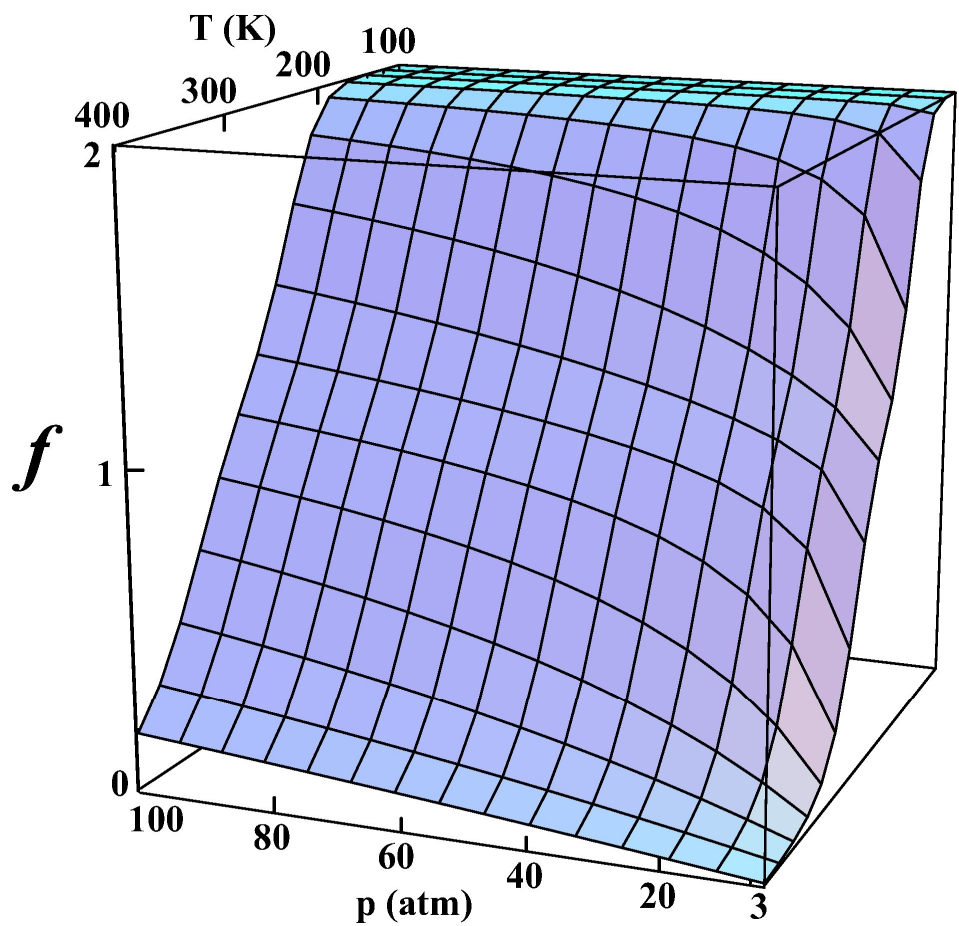

**Figure 3 (Color online):** Occupation number as a function of the pressure and the temperature on the benzyl-released Ti (III) complex.